\documentclass[11pt]{article}
\usepackage[noSecNum]{RBmisc}

\newif\ifTGSA
\TGSAtrue 
\newif\ifarXiv
\arXivtrue 
\newif\ifarXivHide
\ifarXiv{\arXivHidefalse}\else{\arXivHidetrue}\fi
\usepackage{listings}
\newcommand{\citeAll}{\cite{GSEA05,SPIA,SPIAR,draghici2007systems,ROTPE,%
ROTPER,PathNet,NEA,NEAguiR,CePa,CePaR,DEGraph,topologyGSA,topologyGSAR}}

\graphicspath{{FIGS/}}
\begin{document}
\providecommand{\figtitle}{}

\def\keggfigCap{%
\figtitle{Fragment of the KEGG cell-cycle pathway (hsa04110).}
Classical pathway analyses such as GSEA do not take the network topology
into account, but rather treat the pathway as a simple list of genes.
As a result, changes to a gene such as p53 (red), which has a high degree and
a direct influence on a number of other high--degree genes, a large
downstream network, and an outgoing connection to a whole other
network (the apoptosis pathway) are treated in the same way as
changes to a gene such as Bub1 (blue), which has far
fewer connections.  In contrast, topology--based analyses attempt
to incorporate the structure of the network and the relative
importance of each gene to the pathway.
}
\def\keggfigFig{%
	\postfig[width=0.5\textwidth]{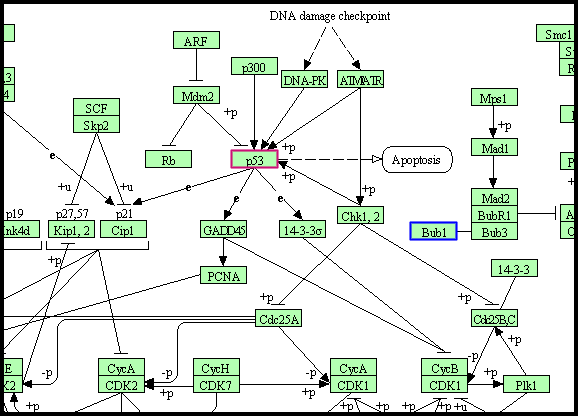}{keggfig}{\keggfigCap}
}

\def\heatPWoverlapCap{%
\figtitle{Number of studies in which a pathway ranks in the top 20\% for each analysis.}
For each subcomputation of each method, we show for each pathway
the number of times the pathway was amongst the top 20\% most
significant in each of the 10 studies.  Pathways were only considered
significant if they met the 20\% cutoff unambiguously; if there
were more than 20\% of pathways tied for the top spot, 
none were considered to be meaningfully in the
top 20\%.  Methods are labeled in alternating colors, with the
final/combined $p$ values denoted in bold.  The number of studies
(out of 10 possible) in which the pathway was in the top 20\% for
that analysis is given by color; black indicates that the method
could not give an answer for that pathway (typically a result of
gene thresholding leaving no meaningful edges). The $p$ values in
the color scale correspond to the probability of that specific
overlap assuming 10 Bernoulli trials with $p=0.2$ success.  The 247
pathways are ordered along the $x$ axis by the mean overlap from
the final (bolded) analyses, while the bottom row shows the average
across all sub-analyses.
}
\def\heatPWoverlapFig{%
  \postfig[width=\textwidth]{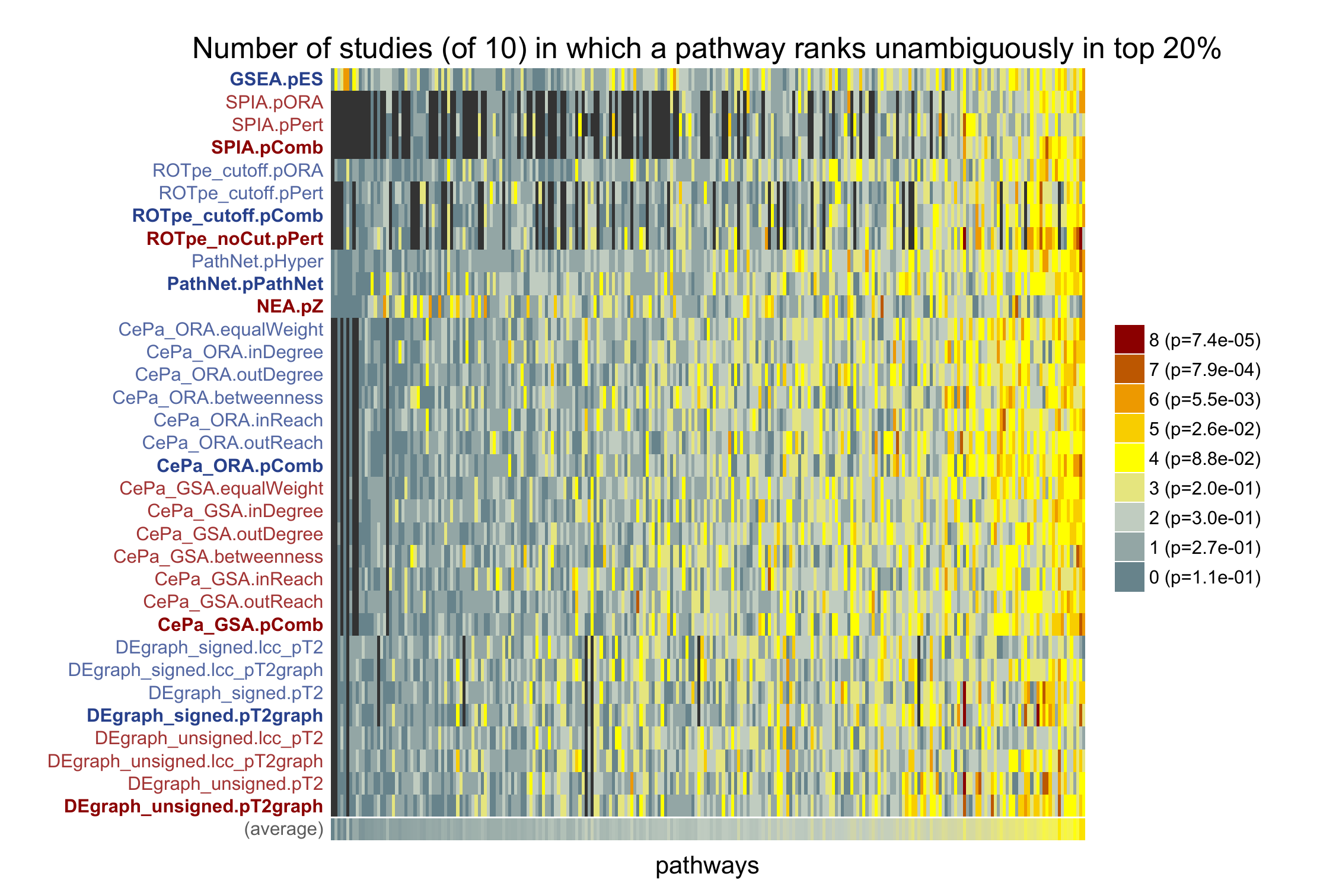}{heatPWoverlap}{\heatPWoverlapCap}
}

\def\windowPaneCap{%
\figtitle{Cross--study concordance for each sub-analysis.}
For each subcomputation of each method, we show the correlation
between pathway $p$ values for all possible study pairs (45 total).
Study pairs are ordered along the $x$ axis according to their
correlation in gene--level $p$ values, shown in the top row.
Methods are labeled in alternating colors, with the
final/combined $p$ values denoted in bold. The bottom row of the
plot shows the sum of the sample sizes for each pair of studies,
with dark green being high.}
\def\windowPaneFig{%
  \postfig[width=\textwidth]{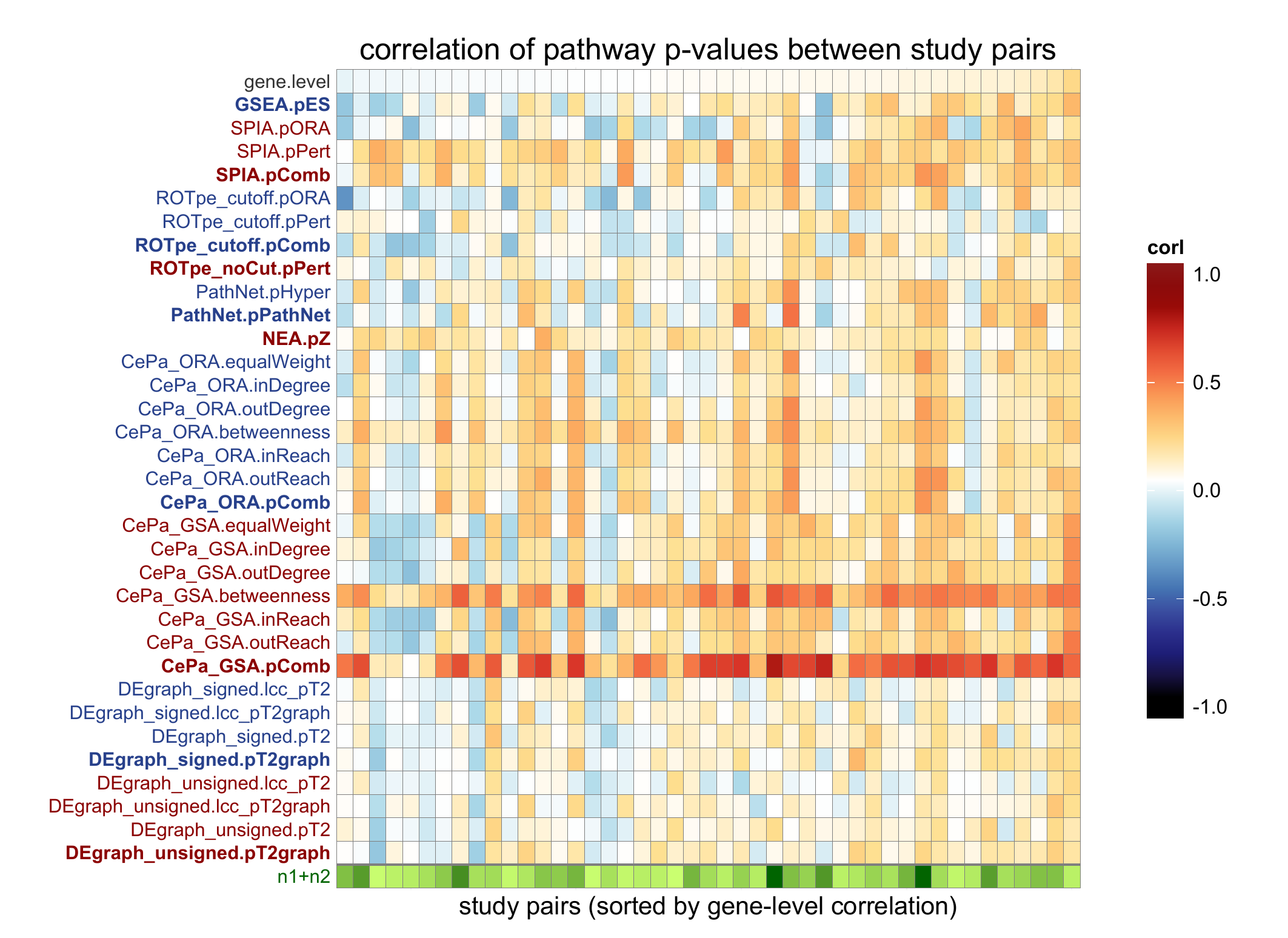}{windowPane}{\windowPaneCap}
}

\def\xStudyCorlCap{%
\figtitle{Cross--study correlations.} 
Each plot displays the cross-study correlation of the results for
each major analysis method.  Boxplots within each frame indicate,
for a given method, 
the distribution of correlations  each study had with the nine other
studies.  EG, consider the ``GSEA.pES'' plot; the blue (leftmost)
box indicates the distribution of correlations of GSEA pathway
enrichment score $p$ values (pES) that study `GSE13876' had with
each of the other data sets.  The red box indicates the correlations
between `GSE14764' GSEA results and those of other nine studies,
etc.  Cross-study correlations of the gene--level statistics are
also shown. Note that the scale on each of the plots is the same.}
\def\xStudyCorlFig{%
  \postfig[width=\textwidth]{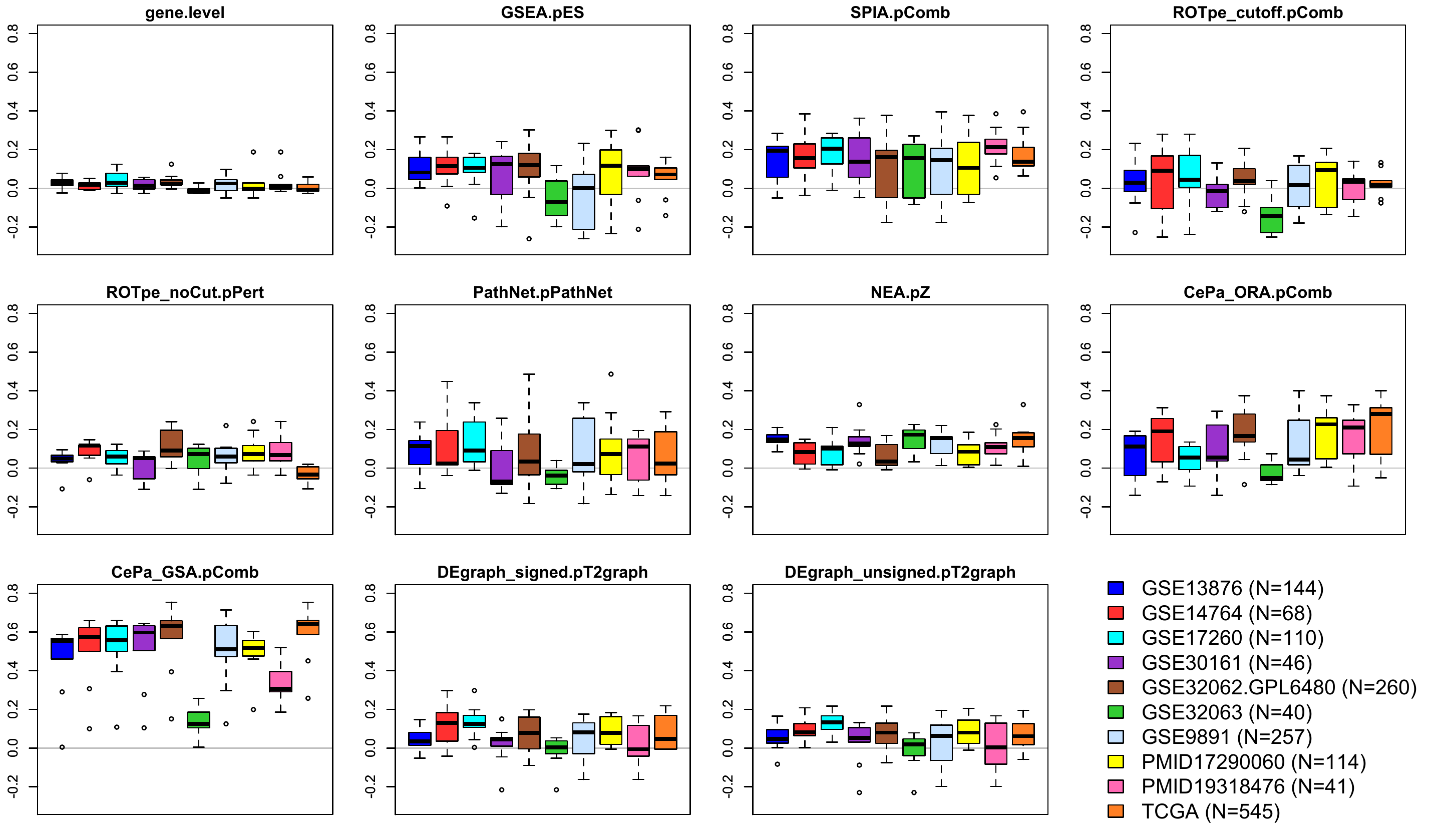}{xStudyCorl}{\xStudyCorlCap}
}

\def\xMethodCorlCap{%
\figtitle{Cross--method correlations.} 
Each plot displays the correlation in $p$ values amongst 
different methods applied to each of the data sets. Here, the
boxplots within each frame indicate, for a given study, the
distribution of correlations the results from each method
had with the others.  EG, in the top left frame, the blue
(left most) box plot indicates the distribution of  correlation
between the pathway enrichment score $p$ values (pES) vs. the
pathway $p$ values obtained from the other nine analyses when
applied to the GSE13876 data.  Note that the scale
on each of the plots is the same.}
\def\xMethodCorlFig{%
  \postfig[width=\textwidth]{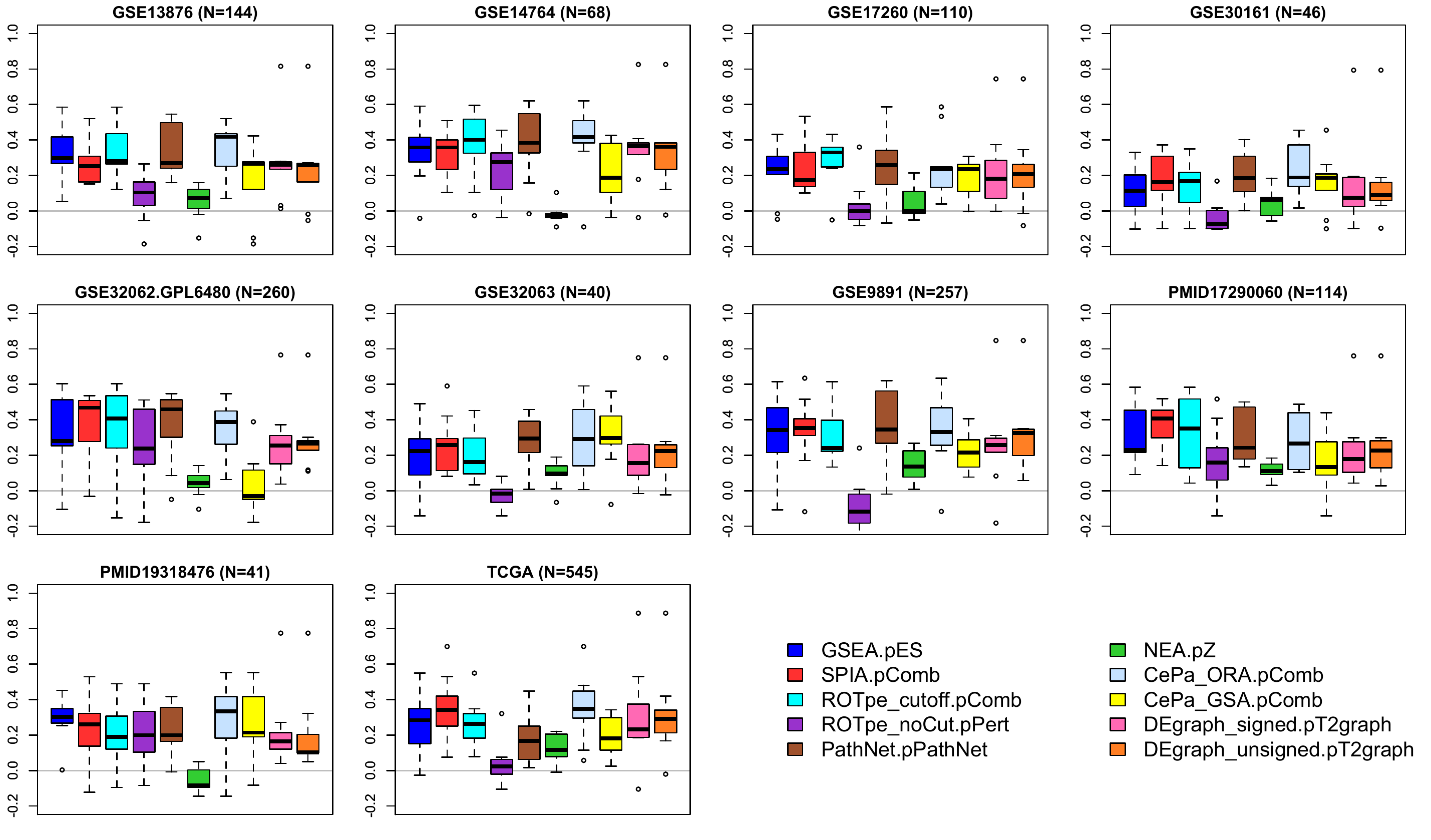}{xMethodCorl}{\xMethodCorlCap}
}

\def\allCorlPairsCap{%
\figtitle{p-value distributions by method (all pathways, all studies).}
Depicted are joint and marginal distributions of $-\log_{10}(p)$
values for all pathways in all studies.  (Note that higher values
are more significant.)  In the upper triangle, smoothed scatter
plots depict the joint distribution of $-\log_{10}(p)$ for each
pair of methods; darker red corresponds to higher density of points.
In the lower triangle, Spearman's \textit{rank} correlations $\rho$ between
the $p$ values obtained from each pair of methods is given, with
positive correlations shown in increasing blue intensity and negative
correlations shown in increasing red intensity (there are no negative
values). Note that because rank correlations provide a measure of concordance
that is independent of the dynamic range of the quantities being correlated
and hence less influenced by outliers, the  $\rho$ reported in the lower 
triangle may differ from a ``by eye'' estimate of the corrlation based on the
plots in the upper triangle. On the diagonal, the marginal distributions
of $-\log_{10}(p)$ are shown as red histograms, with the theoretically 
expected distributions (uniform $p$ under the null) shown as a black line.}
\def\allCorlPairsFig{%
  \postfig[width=\textwidth]{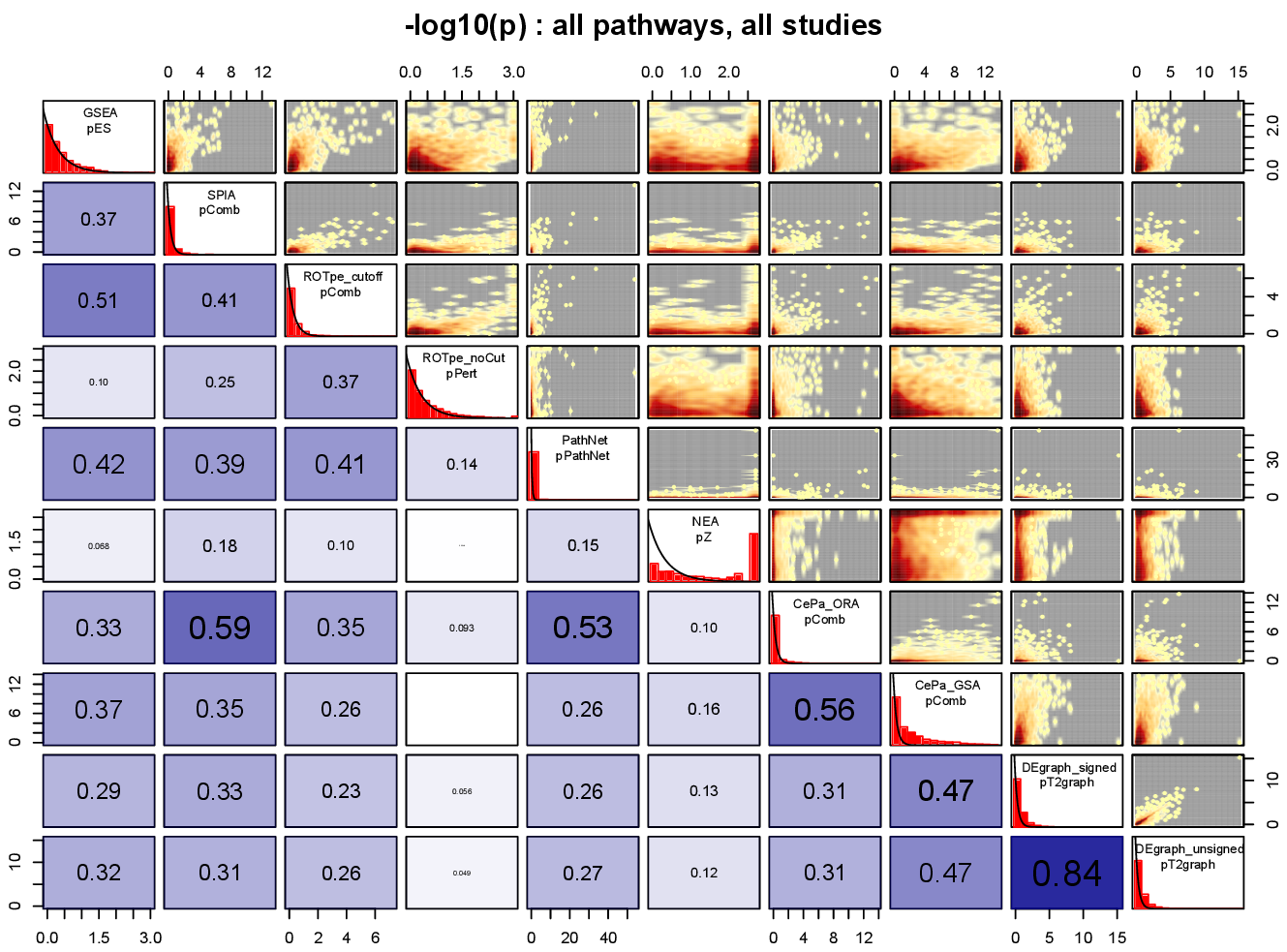}{allCorlPairs}{\allCorlPairsCap}
}

\def\methTab{%
\begin{table}[ht]
\centering
{\footnotesize{
\begin{tabularx}{\textwidth}{llXXXXlr}
\toprule
 & Gene {$p$-value} & Expression$^1$ & 2-sample &  \multicolumn{2}{l}{}  &  & \\
Method &  thresholding & data only & restriction & Directed  & Signed$^2$  & Null type$^3$ &  Citation\\
\midrule
GSEA & no & no & no& N/A & N/A & self-contained&\cite{GSEA05} \\
SPIA & yes & no & no& yes & yes & competitive& \cite{SPIA,SPIAR,draghici2007systems} \\
ROT/pe&optional & no &no& yes & yes & competitive& \cite{draghici2007systems,ROTPE,ROTPER} \\
PathNet&yes& no& no & yes & no & competitive& \cite{PathNet} \\
NEA&yes&no&no&no&no&competitive& \cite{NEA,NEAguiR}\\
CePa-ORA& yes&no&no&yes&no& competitive& \cite{CePa,CePaR}\\
CePa-GSA& no&as written$^4$&as written$^4$&yes&no& self-contained& \cite{CePa,CePaR}\\
DEGraph&no&yes&yes&no&optional&self-contained&\cite{DEGraph}\\
\ifTGSA%
TopologyGSA&no& yes&yes&no&no&self-contained&\cite{topologyGSA,topologyGSAR}\\ \fi
\bottomrule
\end{tabularx}
}}
\caption{\label{tab01} Properies of the various methods, including whether genes are
thresholded on $p$-values, restrictions on the type of data and comparisons that can be
made, the type of edges used by the model, and the type of null hypothesis tested. Notes:\\
$^1$ ``Expression data'' is used here to denote any data that meets the assumptions used
in gene expression testing, i.e., that the data is continuous and normally distributed.  
Other data meeting these assumptions can also be used, but methods which have this restriction
cannot accept SNP data, etc.\\
$^2$ ``Signed'' edges have signs assigned based on the interaction type in the pathway
reference graph, distinguishing activating/inducing edges (+1) from repressing/inhibiting edges (-1).\\
$^3$ ``Null type'' refers to the type of null hypothesis tested
(cf.~\cite{KHAT2012}.  ``Competitive'' null hypotheses compare the pathway
of interest to randomly generated pathways, without permuting sample
labels, \ie, testing that the pathway statistic is more significant than a 
random pathway \textit{given} the sample labels.  ``Self-contained'' null hypotheses
compare the statistic for the pathway to that obtained by randomly permuting
sample labels, \ie, testing that the pathway is more strongly associated
with a particular phenotypic attribute than expectd by chance \textit{given} the 
genes and topology of the pathway.  The self-contained null is considered
to be a stronger test~\cite{KHAT2012}.\\
$^4$ CePa-GSA is not inherently restricted to 2-sample tests of differential expression
methodologically, however
the present implementation will only carry out the non-thresholded ``GSA'' type analysis
for 2-sample tests of differential expression.  As currently implemented, other types
of analyses (\eg, using SNP data or modeling survival) must be carried out in CePa using the
``ORA'' analysis, which requires gene $p$-value thresholding and can only test the competitive
null hypothesis$^3$.
}
\end{table}
}

\def\studTab{%
\begin{table}[ht]
\centering
\begin{tabular}{lrr}
  \toprule
&\multicolumn{2}{c}{grade}\\
\cmidrule(r){2-3}
Study Accession No. & $N$(low) & $N$(high) \\ 
  \midrule
  GSE13876\_eset &  59 &  85 \\ 
  GSE14764\_eset &  24 &  44 \\ 
  GSE17260\_eset &  67 &  43 \\ 
  GSE30161\_eset &  19 &  27 \\ 
  GSE32062.GPL6480\_eset & 131 & 129 \\ 
  GSE32063\_eset &  23 &  17 \\ 
  GSE9891\_eset & 103 & 154 \\ 
  PMID17290060\_eset &  57 &  57 \\ 
  PMID19318476\_eset &  17 &  24 \\ 
  TCGA\_eset &  75 & 470 \\ 
\bottomrule
\end{tabular}
\caption{\label{ovdattab} Studies and samples sizes of the data
used in this investigation.  The data are publicly accessible 
from GEO~\cite{GEO} and 
available as part of the curatedOvarianData package.}
\end{table}
}

\def\timeTab{%
\begin{table}[!hb]
\centering
{\small{
\begin{tabularx}{\textwidth}{llrrrrr@{~}r@{~}X@{}}
\toprule
&&\multicolumn{4}{c}{time (sec), 247 pathways, all 10 studies} &%
\multicolumn{3}{l}{\multirow{2}{*}{\begin{minipage}{13ex}
\flushright avg runtime per study \end{minipage}}} \\
\cmidrule(r){3-6}
Method  &Package Version & user& system&CPU (usr+sys)& wallclock & \multicolumn{3}{l}{} \\
\midrule
SPIA         &\texttt{SPIA\_2.14.0}      & 1229.803& 123.825& 1353.628& 1356.908&   & 2m&15s\\
ROT/pe, cutoff&\texttt{ROntoTools\_1.4.0}&  948.702&   9.497&  958.199&  959.367&   & 1m&35s\\
ROT/pe, no cut&\texttt{ROntoTools\_1.4.0}& 1054.795&  16.861& 1071.656& 1076.913&   & 1m&47s\\
PathNet      &\texttt{PathNet\_1.2.0}    &  747.983&   9.572&  757.555&  796.516&   & 1m&20s\\
NEA          &\texttt{neaGUI\_1.0.0}     &78099.148&5280.723&83379.871&83389.635& ~2h&18m&58s\\
CePa-ORA     &\texttt{CePa\_0.5}         & 3217.099&  21.367& 3238.466& 3238.537&   & 5m&24s\\
CePa-GSA     &\texttt{CePa\_0.5}         &29127.875& 512.618&29640.493&29650.268&   &49m&25s\\
DEGraph (s,u)&\texttt{DEGraph\_1.14.0}   & 2467.075&  43.743& 2510.818& 2514.990&   & 4m&11s\\
\ifTGSA
TopologyGSA &\texttt{topologyGSA\_1.4.3}&\multicolumn{3}{c}{[N/A: job halted after 100hrs]}&${\gg}360000.000$& & & \\ \fi
\bottomrule
\end{tabularx}
}}
\caption{\label{runtab} Package versions and run times (in seconds) for 247
pathways in all ten studies. Average times (in h/m/s) for a single study (\ie, 1/10th
wallclock) are also noted.  Times for computation (``user'') and
kernel system calls (``system'') are given; the total CPU time consumed
is the sum of these.  The computations were carried out on a lightly--loaded 3.4 GHz quad-core
Intel Core i7 iMac with 16GB RAM running R version 3.0.3 under OS X 10.8.5 (Darwin 12.5.0).
Note that the times for DEGraph include both the signed and unsigned graph analyses.
1000 random permutations were used for the methods that perform resampling (all except DEGraph).}
\end{table}
}

\renewcommand\Authands{ and }
\renewcommand\Authfont{\large}
\renewcommand\Affilfont{\small\itshape}

\title{%
Network Methods for Pathway Analysis of Genomic Data
}

\author[1]{Rosemary Braun\thanks{Corresponding author: \texttt{rbraun@northwestern.edu}}}
\author[2]{Sahil Shah}

\affil[1]{Biostatistics, Feinberg School of Medicine and Northwestern Institute on Complex Systems}
\affil[2]{Engineering Sciences and Applied Mathematics}
\affil[ ]{Northwestern University}

\date{}

\maketitle

\begin{abstract}
Rapid advances in high--throughput technologies have led to
considerable interest in analyzing genome--scale data in the context
of biological pathways, with the goal of identifying functional
systems that are involved in a given phenotype. In the most common
approaches, biological pathways are modeled as simple sets of genes,
neglecting the network of interactions comprising the pathway and
treating all genes as equally important to the pathway's function.
Recently, a number of new methods have been proposed to integrate
pathway topology in the analyses, harnessing existing knowledge and
enabling more nuanced models of complex biological systems.  However,
there is little guidance available to researches choosing between
these methods.  In this review, we discuss eight topology-based
methods, comparing their methodological approaches and appropriate
use cases.  In addition, we present the results of the application
of these methods to a curated set of ten gene expression profiling
studies using a common set of pathway annotations.  We report the
computational efficiency of the methods and the consistency of the
results across methods and studies to help guide users in choosing
a method.  We also discuss the challenges and future outlook for
improved network analysis methodologies.

\end{abstract}
\section{Introduction}
Modern high--throughput (HT) technologies enable researchers to make
comprehensive measurements of the molecular state of biological
samples and have yielded a wealth of information regarding the 
association of genes with specific phenotypes.  However, the complex
and adaptive nature of living systems presents a significant challenge
to deriving accurate and predictive mechanistic models from genomic
data.  Because cellular processes are governed by networks of 
molecular interactions, critical alterations to these systems may
arise at different points yet result in similar phenotypes.
At the same time, the adaptability and robustness of living
systems enables variations to be  tolerated. Typical gene--level
analyses of HT data, such as tests of differential expression, are
unable capture these effects.  As a result, there has been growing
interest in systems--level analyses of genomic data.

Pathway analysis techniques, which aim to examine HT data in the
context of mechanistically related gene sets, have been enabled by 
the growth of databases describing functional networks of 
interactions.  These include KEGG~\cite{KEGG}, BioCarta~\cite{BioCarta},
Reactome~\cite{reactome}, the NCI Pathway Interaction Database
(NCI-PID)~\cite{PID}, and InnateDB~\cite{lynn2008innatedb}, amongst others.
To address the challenge of querying these databases using a common
framework, markup languages such as KGML (used by KEGG) and BioPAX
have been developed to describe pathways using a consistent format.
In particular, the Biological Pathway Exchange (BioPAX) project now
provides a unified view of the data from many of the above
sources~\cite{BioPAX}, including NCI-PID, Reactome, BioCarta, and
WikiPathways.

Over the past decade, a number of pathway analysis methods have 
been developed to integrate this information with data derived
from genomic studies~\cite{KHAT2012}. These methods can be broadly
grouped into two categories.  The first category comprises analyses designed
to identify pathways in which significant genes are overrepresented.  A
comprehensive review of these methods was recently published Khatri
\&al~\cite{KHAT2012}; examples include hypergeometric tests, Gene
Set Enrichment Analysis (GSEA)~\cite{GSEA05}, and Ingenuity Pathway
Analysis~\cite{Ingenuity}.  The second category of approaches use
dimension reduction algorithms to summarize the variation of the
genes across the pathway and test for pathway--level differences
without relying on single--gene association statistics.  
Examples include GPC-Score~\cite{BRAU2008}, Pathifier~\cite{DRIE2013},
and PDM~\cite{BRAU2013}.  In contrast to enrichment analyses like
GSEA, these methods are capable of identifying differences at the
systems--level that would be indetectable by methods that rely
upon single--gene association statistics, such as differences in
the \textit{coordination} of the expression of two genes.  

Despite these advances, the majority of these methods treat pathways
as simple lists of genes, neglecting the network of interactions
codified in pathway databases (cf. Fig~\ref{keggfig}) despite the
fact that the importance of pathway network structure to biological
function has long been appreciated.
In~\cite{JEON00}, the authors presented systematic mathematical
analysis of the topology  of metabolic networks of 43 organisms
representing all three domains of life, and found that despite
significant variation in the pathway components, these networks
share common mathematical properties which enhance error-tolerance.
In~\cite{JEON01}, the authors compared the lethality of mutations
in yeast with the positions of the affected protein in known pathways,
and found that the biological necessity of the protein was well
modeled by its connectivity in the network.

To incorporate known interaction network topology with traditional
pathway analyses, a multitude of approaches have been proposed for
overlaying gene--specific data (either the raw data itself or
$p$-values derived from gene--level statistical tests) onto pathway
networks~\cite{%
SPIA,
draghici2007systems,ROTPE,ROTPER,
PathNet,
NEA,NEAguiR,
CePa,CePaR,
DEGraph,
topologyGSA,topologyGSAR,
shojaie2010penalized,dittrich2008identifying,beisser2010bionet,IDEK01,BRAU2012,EFRO07,PathOlogist,NetGSA}.
However, to date few comparisons between them have been made.  One
recent review~\cite{VARA2012} attempted to compare three network--based
analyses (SPIA, PARADIGM, and PathOlogist); unfortunately, the
comparison was stymied by the methods' disparate implementations
and reference databases, and yielded inconclusive results.
To address this gap, we review eight topology--based pathway analysis
methods~\citeAll.  While this is not a comprehensive review of all
such algorithms, the methods we consider have the common feature
of being implemented in R and permitting the user to provide the
pathway models (such as those obtained from KEGG), thus allowing
them to be compared directly without the issues encountered
in~\cite{VARA2012}.  Their free availability from CRAN and
BioConductor~\cite{gentleman2004bioconductor} also makes them the
most popular network analysis methods in the bioinformatics and
computational biology research community.  

Below, we briefly describe each method and discuss its features and
limitations.  In addition, we also provide a comparison of these
methods applied to a curated set of gene expression data from 10
ovarian cancer studies~\cite{GANZ2013}, using a common set of 247
KEGG pathways.  Using this data, we were able to evaluate both the
computational efficiency of the methods and the consistency of the
results between the methods and across the studies.  The suite of
data and scripts used in this comparison are available from our
website, enabling researchers to compare updated versions and new
network analysis packages using this common framework.

\subsection{Overview of Network Pathway Analysis Methods}
We consider 8 popular network-based pathway analysis methods,
described below.  In the following discussion, we consider a pathway
to be a network of genes (nodes), where edges represent a biochemical
interaction.  These edges may be directed (\eg, gene $i$ induces
gene $j$, but not vice-versa) or signed (\eg, $+1$ for activation
or induction, $-1$ for repression or inhibition). Also, while the
following methods were designed with differential expression analyses
in mind, it should be noted that most are flexible enough to
accommodate other statistics or data types (\eg, allele frequencies
and $\chi^2$ statistics from GWAS).  A summary of the key features
of the methods is given in Table~\ref{tab01}.  For comparison, we
also consider GSEA~\cite{GSEA05}, a commonly--used pathway analysis
algorithm that does not incorporate network topology.

\paragraph{GSEA}
Gene Set Enrichment Analysis (GSEA)~\cite{GSEA05} is an extremely
popular method for detecting pathways that are enriched for 
differentially expressed genes.  In contrast to simple hypergeometric
overrepresentation analysis, GSEA does not require a threshold for
calling a gene significant, but rather considers the magnitude of
the expression changes for all the genes in the pathway. Briefly,
the algorithm ranks all assayed genes according to the significance of 
the gene--level associations, and computes a running sum statistic
to test whether the genes on the pathway of interest tend to lie
at the top of the ranked list (vs. a null hypothesis of being 
randomly distributed).  Significant pathways are those whose genes
occur higher in the ranked list than expected by chance, resulting
in a sharply peaked running sum statistic.  The size of this peak,
called the enrichment score (ES), is tested for significance by
permuting the sample labels and recalculating the test statistic
to generate a null distribution.  This procedure identifies pathways
with an accumulation of differentially expressed genes that are
associated with the phenotypes of interest. As its name suggests,
however, GSEA treats pathways as \textit{sets} of genes, without
incorporating network topology.

\paragraph{SPIA}
Signaling Pathway Impact Analysis (SPIA)~\cite{SPIA}, and the
related ROntoTools/Pathway-Express method described below,
are methods designed to quantify the impact that differentially
expressed (DE) genes have on the downstream elements of the 
pathway by taking into account the number of DE genes, the
magnitude of the expression differences, and the direction
and type of edges in the pathway.  In SPIA, a ``perturbation
factor'' is computed for each gene $g$, defined as the weighted
sum of the expression differences of all the DE genes which
are direct upstream parents of $g$ in the network.  The weighting
coefficients are a function of both the number of outgoing edges of
the parent node, as well as the type of edge (positive for
induction/activation, negative for repression/inhibition) connecting
the parent to the child node $g$.   The total accumulated perturbations
of all genes in the pathway is compared against a null distribution
obtained by repeatedly randomizing the observed expression differences
to different genes in the network.  SPIA combines the resulting
perturbation $p$ value, $p.$PERT, with the $p$ value from the simple
hypergeometric overrepresentation test for the pathway, $p.$ORA,
using Stouffer's weighted-$z$ method~\cite{Stouffer}, yielding a
pathway graph $p$ value ($p.$Comb).  Thus, a pathway with DE genes
at its entry point could be deemed more significant than one with
DE genes further downstream, even if the fold changes are smaller.

Although SPIA is described in the context of differential gene
expression, the method can in principle be used with any type of
gene-level association statistic (\eg, hazard ratios for survival,
minor allele frequencies of SNP variants, or multivariate $F$ or
$T^2$ statistics testing association across more than two phenotypes).
However, SPIA requires that the user set a significance threshold 
to select the genes that are considered in the analysis.

\paragraph{ROntoTools/Pathway Express}
ROntoTools/Pathway Express (ROT/pe)~\cite{draghici2007systems,ROTPE,ROTPER}
is a refinement upon SPIA.  In SPIA, a gene significance threshold
is used to select DE genes, and the DE genes are treated equally,
while those that are not DE are not considered in the analysis.  To
enable a more nuanced analysis, ROT/pe weights
the change in expression of each gene by the significance of the
gene level statistics. As a result, marginally significant
changes in expression are given less weight than a highly significant
changes of the same magnitude.  Furthermore, this method permits a
``cut-off free'' analysis that assigns non-significant genes a low
weight rather than aggressively discarding them as non-DE.  When
no cutoff is used, ROT/pe does not calculate the overrepresentation
significance $p.$ORA, but simply reports the significance of the
perturbation (obtained, as in SPIA, by randomly permuting the gene
expression differences across the network).

\paragraph{PathNet}
PathNet~\cite{PathNet} incorporates pathway topology by computing
both a ``direct'' and ``indirect''  association statistic for each
gene.  The ``direct'' statistic is the result of a single--gene
association test (\eg, a classical $t$ test of differential expression).
Once these are obtained, PathNet then computes ``indirect'' statistic
for each gene, defined as the sum of the $-\log_{10}{p}$ values of
the direct associations for all the gene's neighbors in the pathway.
The direct and indirect scores for each gene are summed and tested
for significance by permuting the direct evidence statistic across
all genes in a ``global'' network formed by merging pathways with
common genes.  The resulting gene--level $p$ values are then
thresholded for significance, and the pathway is tested for enrichment
using a hypergeometric test.

Like SPIA and ROT/pe, PathNet can accommodate a variety of gene--level
metrics and is not restricted to simple differential expression.
In contrast to SPIA and ROT/pe, PathNet considers all possible edges
in the global network, not simply those confined to a single pathway,
enabling it to find pathways that are strongly affected through
indirect links. Also in contrast to both SPIA and ROT/pe, PathNet
does not distinguish between inhibitory and activating edges.
PathNet's use of thresholding for testing pathway significance is
also a drawback, but could easily be overcome by using a GSEA-like
procedure instead of the hypergeometric test.

\paragraph{NEA}
As in PathNet, Network Enrichment Analysis (NEA)~\cite{NEA,NEAguiR} also
attempts to incorporate information from the ``global'' network of
pathways formed by merging the common genes in the individual
functional pathways represented in the database, but uses a 
different approach to quantify enrichment.  Rather than counting 
significant nodes (\ie genes) in each pathway of interest, NEA counts the 
number of edges in the pathway connecting to a significance gene.
This is done by summing the degree (\ie, number of edges) for each
DE gene on the pathway and amongst its neighbors, subtracting out links
from genes that do not connect to a gene in the pathway.  Genes outside
the pathway will thus contribute to the pathway's score; connections
between two DE genes on the pathway are counted twice (once per each gene).
NEA assess the significance of this statistic by randomly rewiring the 
topology of the ``global'' network, conditioned on preserving the 
degree of each node.  

Like all the foregoing methods, any gene-level statistic of interest
may be used as an input to NEA.  However, NEA as implemented is dependent
upon setting cutoffs for gene significance in order to count the
number of DE links, making the results susceptible to noise induced
by thresholding.  It may be of interest to investigate  possible
refinements of NEA using a weighting scheme rather than a strict
binary cutoff.  Also, NEA as currently implemented does not consider
the direction of the edges in the graph, but rather treats all edges 
as bi-directional.  Direction-based refinements to NEA may thus
also be of interest.

\paragraph{CePa}
All of the above methods improve upon GSEA by not only considering
each gene's differential expression, but also examining the
differential expression of its nearest--neighbors (``perturbations''
from upstream genes in SPIA and ROT/pe, ``indirect evidence'' in
PathNet, number of DE links in NEA).  The resulting pathway scores
thus represent an accumulation of \textit{localized} effects within
the pathway.  In contrast to these approaches, CePa~\cite{CePa,CePaR}
attempts to take a broader view of the network by incorporating
graph centrality measures into the statistics.

Briefly, the ``centrality'' of a node in a graph is a measurement
of its relative importance to the rest of the network~\cite{WEST01}.
One simple centrality measure is the node's degree: the number of
edges connecting to that node.  On a directed graph, the
in--/out--degrees, which are the counts of a node's incoming/outgoing
edges respectively, are also useful for quantifying how
susceptible/influential a node is.  ``Betweenness,'' another
centrality measure, quantifies how frequently the shortest path
between any two nodes goes through the node of interest.

CePa uses the centrality measures  of the genes in the network to
perform either an ``over representation analysis'' (CePa-ORA) or
a ``gene set analysis'' (CePa-GSA).   In CePa-ORA, the user specifies
a set of significant genes (based on some pre-determined significance
threshold); CePa-ORA then sums the centrality measures of the significant
genes and calculates the significance of the sum by randomly selecting
a new set of ``significant'' genes based on the proportion of truly
observed significant genes in the network.  That is, CePa-ORA tests
whether significant differential expression is more likely to be
high--centrality nodes in the pathway than would be expected by
chance.

The CePa--GSA variant incorporates the gene--level statistics
(such as the $t$ statistic from a test of differential expression)
rather than using only the list of significant genes.  CePa-GSA
multiplies the gene statistics by the gene centrality measures.  In
this way, CePa directly weights each gene's statistic by its
importance in the network, reflecting the observation made by Jeong
and others~\cite{JEON01} that alterations to more central genes
have a more profound impact on an organism.  CePa--GSA then aggregates
the weighted gene--level statistics into a single pathway--level
statistic by taking, for example, their maximum or median.  As
implemented in the CePa package, the default gene--level statistic
is the absolute value of the $t$ statistic from a two--sample test
of differential expression, and the default pathway--level statistic
is the mean of the centrality--weighted gene--level statistics.
The significance of this pathway--level score is calculated by
permuting the sample labels of the the gene expression matrix,
recomputing the gene--level statistics with the permuted samples,
and then recomputing the pathway statistic.  The resulting $p$ value
quantifies the differences in centrality--weighted gene expression
that are associated with the phenotypic differences of interest.

While in principle both CePa-ORA and CePa-GSA can use any gene--level
statistics, at present the implementation of CePa-GSA is such that
the user is confined only to two--sample $t$--tests of differential
expression.  If, for instance, the outcome of interest is survival
(yielding hazard ratios for each gene), or if differences across
multiple conditions or phenotypes are being assessed with a
multivariate linear model or ANOVA, or if the input data is categorical
rather than continuous (such as allele frequencies from SNP GWAS
studies), the user has no choice other than to precompute the
gene--level $p$ values and use the thresholded CePa-ORA analysis
rather than the more nuanced CePa-GSA approach.  It would be of
considerable interest to address this limitation by providing a
more flexible interface to CePa-GSA that permits user--defined
models.  In addition, while CePa provides a very rich view of the
centrality--weighted pathway analysis, the variety of statistics
it obtains (one per each of 6 centrality metrics considered) and
the diverse options for pathway aggregation  (max, mean, etc., as
well as the option for user--defined function) can make the results
difficult to interpret.  Finally, it may also be argued that 
linear centrality--weighting (as opposed to, \eg, weighing by a
power function of the centrality) is an arbitrary choice that may 
influence the results.

\paragraph{DEGraph}
An alternative topology--weighting approach is implemented in
DEGraph~\cite{DEGraph}, which uses a multivariate Hotelling 
$T^2$ test to identify pathways in which a significant 
subset of genes are differentially expressed.  In DEGraph,
both the standard $T^2$ and a network--smoothed graph $T^2$
are computed.  Based on the intuition that two genes connected
in the network should behave in a correlated fashion, DEGraph's
network--smoothing filters the gene expression differences
by keeping only the first few components of its projection onto
a basis defined by the graph Laplacian~\cite{CHUN97}.  
Spectral decomposition of the pathway's graph Laplacian encapsulates
the geometry of the network topology at progressively finer scales,
and the projection of the gene expression shifts onto the coarsest
components may be analogized to keeping only the lowest frequency
Fourier components of a function.  These network--smoothed shifts
are then tested using the Hotelling $T^2$ test.

Unlike CePa, DEGraph provides a means to obtain pathway--wide
topology--weighted scores without making arbitrary choices about
the weighting.  In particular, the mathematical and statistical
properties of graph Laplacian spectra are
well--characterized~\cite{CHUN97,NG2002} and can be precisely
related to the dynamical properties of the network~\cite{ATAY06,LU07},
making this approach more justifiable than the weighting scheme proposed 
in CePa.  However, this spectral approach is necessarily confined to
operate only on connected components, posing problems for pathways
comprising several disjoint subgraphs (\eg, two sub-processes
connected by a exogenous stimulus that is not represented in the
gene network).  In such cases, the pathway will have several 
$T^2$ statistics---one for each connected component---leaving a choice
as to whether the pathway $p$ value should be defined by that of its
largest subnetwork or some combination of all its subnetworks.
In addition, the use of the Hotelling $T^2$ test requires that the
gene--level statistics follow a multivariate normal distribution,
which limits the types of data and analyses to which DEGraph could
be applied.  In consequence, the DEGraph implementation is only
capable of testing pathway--wide multi--gene differential expression
between two sample classes.  
Recently, a more flexible spectral decomposition method has been
proposed~\cite{BRAU2012,BRAU2010}, but which relies upon computationally
intensive permutation tests.
It would be highly interesting to extend DEGraph  to accommodate
other data types and statistical tests.  

\ifTGSA
\paragraph{Topology GSA}
Although analyzing the data at the pathway level can identify
important systems--level changes that would be missed by single--gene
analyses, the pathway--level findings can be difficult to interpret
and validate for large networks.  While identifying functionally
significant pathways is the goal of systems biology, it is often
necessary to identify specific features within those networks that
can be targetted experimentally.  To facilitate the discovery of
targetable sets of genes and interactions, a number of techniques
have been proposed to search for significant submodules within
pathways.  One such method is TopologyGSA~\cite{topologyGSA,topologyGSAR},
which uses a Gaussian network model to identify significant subnetworks
in the graph.  TopologyGSA begins by transforming the directed
pathway network graph into its so--called ``moral'' graph by
connecting all ``parent'' nodes of a vertex and removing the edge
directionality.  The moral graph is then decomposed into cliques:
subsets of nodes in the graph for which every pair is connected by
an edge (triangles are the simplest cliques; a clique of four nodes
has 6 edges; a clique of five has 10; etc).  Each clique is then
tested for differential expression by modeling the expression of
the genes in the network as Gaussian random variables, subject to
the class--conditional covariance between the clique's genes for
each phenotype.  If no significant cliques are found, edges are
iteratively added and the statistics recomputed.  The result is the
identification not only of pathways with significant gene expression
differences, but of specific connected subgraphs within the pathways
that can then be investigated in greater detail.

An appealing feature of TopologyGSA is the use of gene covariance
in the analysis.  Despite the fact that genes often exhibit
considerable correlations in expression, most pathway analyses
consider the genes as independent variables.  However, TopologyGSA's
approach of conditioning the clique statistics on phenotype--specific
covariances also introduces a limitation: like DEGraph, it can only
be used if the outcome of interest can be dichotomized.

Moreover, while testing cliques of a triangulated graph to detect
significant subnetworks is an interesting idea, it has a very serious
drawback: finding all maximal cliques in a graph is an $NP$-hard
problem, meaning that a solution is not guaranteed in polynomial
time.  In particular, a brute-force search for a clique of size $k$
in a graph with $n$ nodes has a computational complexity of
$\bigO(n^kk^2)$.  While approximations can be made, a recent proof
by Chen~\cite{CHEN06} demonstrated that the clique problem cannot
be solved in less than $f(k)n^{o(k)}$ time for \textit{any} function
$f$ and linear function $o(k)$.  As the size of the pathway $n$
grows,  the size of the submodule cliques $k$ also increases, and
time needed to search for those submodules increases exponentially.
This means that for larger pathways---often the very ones for which
submodule detection is desirable---the problem TopologyGSA attempts
to solve may be intractable.  Because TopologyGSA iteratively adds
edges as part of the significance computation, the exponential cost
is incurred on each iteration.  A far more efficient approach would
be to use spectral methods to identify communities of
nodes~\cite{CHUN97,NG2002,Newman:2004p2491,Danon:2005p2497} rather
than articulating all maximal cliques. Such an approach, which would
be closely related to yet still distinct from the spectral used in
DEGraph~\cite{DEGraph} and Pathway/PDM~\cite{BRAU2012,BRAU2010},
would permit the identification of connected subnets of significant
genes far more efficiently than solving the clique problem.  
\fi

\subsection{Evaluating the Performance of Network Analysis Methods}
As described above and shown in Table~\ref{tab01}, these network
analysis methods have different features that make them better
suited to some use cases than others.  Nevertheless, for many common
analyses, most of these approaches could be applied, and the
user is faced with a choice between several promising methods.
Unfortunately, benchmark tests to systematically evaluate the
performance of network analysis methods remain lacking, limiting
the community's ability to compare methods.  

The development of a systematic evaluation framework faces a number
of challenges.  First, the methods themselves have highly disparate
implementations, often using different databases and pathway
semantics, making them difficult to compare in a consistent way.
For example, PathOlogist~\cite{PathOlogist,EFRO07} treats pathways
as bipartite graphs of genes and interactions in contrast to the
gene--mode networks considered by the methods described above, and
is restricted in its implementation to pathways from the
NCI/PID~\cite{PID} database.  Secondly, it is not clear what the
``gold standard'' for these methods should be.  Unlike machine
learning and  network inference problems which are readily tested
against simulated benchmark data with known solutions (such as the
in-silico data suites used in the DREAM
Challenges~\cite{marbach2010revealing}), there is no agreement on
what the ``correct'' results of these analyses would be.

To provide intuition regarding the performance of the methods
reviewed here, we systematically applied them to a curated suite
of gene expression data from 10 ovarian cancer studies~\cite{GANZ2013}
using a common set of pathway definitions obtained from the KEGG
database.  The goal of these tests was not only to supply provisional
guidance about the relative performance of the methods, but also to
suggest a strategy for a testing framework for pathway analysis
methods.

\section{Methods}
Our approach is motivated by the observation that systems--level
analyses improve the concordance of results between different studies
of the same phenotype~\cite{BRAU2010,MANO06}.  Although multiple
studies of the same phenotype may yield very different lists of
significant genes, pathway analyses tend to to show much greater
agreement.  This effect is not unexpected, considering the complexity
of biological systems~\cite{HANA00} and the noisiness of HT data;
individual disease--associated genes may be detected in some studies
but miss the significance threshold in others.  However, if a pathway
is functionally related to the disease, we may reasonably expect to
detect its association across multiple studies, even if the specific
genes contributing to its significance vary from one study to the
next.

This observation leads to the following conjectures:  If a specific
pathway is functionally related to a particular phenotype, we expect
that some  manifestation of its involvement will be present in the
data for all studies of that disease, and that an accurate and
sensitive network analysis approach will detect those signals
consistently across the studies.  A poor network analysis method,
on the other hand, will yield results that are strongly influenced
by noise in the data, and hence will detect pathways that are
particular to each study rather than the common biological signal.
On the basis of this conjecture, we use the cross--study concordance
of each method's results to measure its ability to detect a common
(and presumably ``true'') signal in each of the studies.

\subsection{Ovarian Cancer Data}
For the purposes of our analysis, we used gene expression and
clinical information from curatedOvarianData~\cite{GANZ2013}, an
expert--curated collection of uniformly prepared microarray data
and documented clinical metadata from 23 ovarian cancer studies
totalling 2970 patients.  The curatedOvarianData project was designed
to facilitate gene expression meta-analysis as well as software
development.  By providing a consistent representation of data that
has been processed to ensure comparability between studies, the
package enables users to immediately analyze the data without needing
to reconcile different microarray technologies, study designs,
expression preprocessing methods, or clinical data formats.

Because several of the methods under consideration were limited to
two--sample comparisons, we selected data sets with sample
classes that could be meaningfully dichotomized.  Since the vast
majority of the samples came from patients with stage III cancers,
tumor/normal and stage--based comparisons were not feasible; instead,
we chose to compare low-- and high--grade ovarian serous carcinomas.
These grades have distinct histological features, molecular
characteristics, and clinical outcomes~\cite{singer2003mutations,VANG2009}.
Low--grade serous carcinomas typically evolve slowly from adenofibromas,
acquiring over time frequent mutations to KRAS, BRAF, or ERBB2
genes, but not TP53 mutations.  In contrast, high--grade serous
carcinomas are characterized by TP53 mutations, often without
mutations to KRAS, BRAF, or ERBB2.  They arise from unknown precursor
lesions, progress rapidly, and have worse clinical outcomes.  For
this analysis, we selected studies with a minimum of 15 high-- and
low--grade serous carcinomas and 1000 genes assayed, keeping only
the patients who fell into those categories and who had survival
data.  10 of the 23 available studies met these criteria.  The study
accession numbers and sample counts are given in Table~\ref{ovdattab}.

The microarray data was filtered to keep only the genes common to
all 10 studies; no other filtering was done.  This resulted in 7680
genes common to all 10 studies.

To obtain the gene--level statistics required by several of the
analyses, the R limma package~\cite{smyth2005limma} was used.
$\log_2$ fold changes were used for the magnitude of differential
expression when required; the significance of the association was
quantified using the $p$ value for the empirical Bayes estimated
$t$ statistic~\cite{smyth2005limma}.  Where thresholds for significance
were needed, the  $0.05$ most significant genes were selected. (NB,
this corresponds to the $0.05$ quantile of significance, \textit{not}
$p=0.05$.  Because the studies varied considerably in their sample
sizes, and hence power, we chose to use a quantile--based threshold
rather than a $p$ value threshold to render them comparable.  While
the $p$ value for the $0.05$ quantile varied from study to study,
in all cases this corresponded to $p\ll0.05$.)

\subsection{Network Models}
In order to ensure that each of the eight methods tested used a
common, comparable set of pathway definitions, we created the pathway
annotation objects required for each method by hand from a fresh
download of the KEGG pathway database~\cite{KEGG}. The KEGGgraph R
package~\cite{KEGGgraph} was used to obtain the pathway KGML files
for 247 human pathways.  The KGML files were first processed into
R \verb|graphNEL| objects for use by ROntoTools/PathwayExpress
(ROT/pe), \ifTGSA{TopologyGSA,}\fi and DEGraph.  The \verb|graphNEL|
objects were then used to to generate the lists of genes, edges,
and adjacency matrices variously required by GSEA, PathNet, NEA,
and CePa.  The \verb|path.info| data used by SPIA was also generated
from the KEGG \verb|graphNEL| objects and written to disk as required
by SPIA.  In this way, we ensured that the set of pathways considered
by each method would be directly comparable to each other.  The
complete set of pathway annotation objects for all the methods,
along with an R script to generate a complete set of updated mappings
from a fresh KEGG download, is available from
\url{http://braun.tx0.org/netRev}.

\subsection{Application}
Each of the methods shown in Table~\ref{tab01} was applied to all
10 data sets described in Table~\ref{ovdattab} for all 247 KEGG
pathways in out database.   Where
permutation tests were required, 1000 permutations were used. 
Several of the methods have options that permit different styles
of analysis, which we also explored.  The details of our application
is given below:
\begin{description}
\item[GSEA]~\\
As a point of reference, the non--networked GSEA 
was applied as described in~\cite{GSEA}, using
the gene $p$-values obtained from limma as described above. 
Significance was tested using 1000 permutations of the sample
classes.
\item[SPIA]~\\ 
SPIA was applied using an 0.05--quantile threshold
for significance as described above.  A number of the pathways
had no edges considered by SPIA (which only considers directed
edges, cf.~\cite{SPIA,SPIAR}), and so were
preemptively excluded from the analysis by the package. 
The overrepresentation and perturbation $p$-values were combined
using Stouffer's normal--inverse method.
\item[ROT/pe, cutoff]~\\ 
ROT/pe permits analysis with and without a $p$ value
threshold~\cite{draghici2007systems,ROTPE,ROTPER}; we applied both.
Here, we used the same threshold as in SPIA, meaning that the results
should be roughly comparable to SPIA for the $p.$ORA overrepresentation
analysis.  In contrast to SPIA, however, pPert is now weighted by
the gene's significance, rather than treating all significant genes
equally.  Genes not meeting the significance threshold are excluded
from $p.$Pert with 0 weight, similar to the exclusion in SPIA.
$p.$Pert and $p.$ORA were combined as in SPIA.
\item[ROT/pe, no cut]~\\ 
We performed ROT/pe without a significance
threshold.  Because the hypergeometric test cannot be performed
without setting a cut-off, only pPert is reported.  In contrast
to the thresholded analysis, pPert now involves data from 
\textit{all} the genes, although those with low significance will
have low weighing.
\item[PathNet]~\\ 
PathNet was carried out as described above and
in~\cite{PathNet}.  PathNet returns both the PathNet $p$ value
combining the ``direct'' and ``indirect'' evidence, along with
the simple hypergeometric $p$ value.  The quantity of interest
is $p$PathNet.
\item[NEA]~\\ 
NEA was carried out as described above and in~\cite{NEA,NEAguiR},
using the same gene thresholds and number of permutations as in
the other studies.
\item[CePa-ORA]~\\
Like ROT/pe, CePa also has options to perform the analyses with
or without setting a gene--significance threshold~\cite{CePa,CePaR}.  
We performed both; here, we use the same thresholds used in the
other analyses to carry out CePa-ORA.  As described above, CePa
will report the significance using a variety of network centrality
measures.  Because there is no clear choice of which one is correct,
we chose to combine all six measures into a single $p$ value for
CePa-ORA using Stouffer's method.
\item[CePa-GSA]~\\
We also performed the non-thresholded CePa-GSA. The analysis
differs from CePa-ORA not only in the number of genes considered,
but also in the type of hypothesis test performed.  While CePa-ORA
tests a ``competitive'' hypothesis~\cite{KHAT2012} comparing pathways to random
subsets of genes while holding fixed the sample labels (and hence
the gene--level statistics), CePa-GSA tests the ``self--contained''
null hypothesis by permuting the sample labels while holding the
pathway definition fixed (see Table~\ref{tab01}).  The two tests are
orthogonal to each other, and we do not anticipate that the 
results of CePa-GSA will necessarily be the same as those for CePa-ORA.
As in CePa-ORA, we chose to combine the six $p$'s into one
$p$ value for CePa-GSA using Stouffer's method.
\item[DEGraph]~\\
DEGraph also presents several alternative analysis approaches,
specifically, whether or not the network should be signed
(corresponding to inversely--related nodes) or unsigned for the
purposes of computing the smoothing vector~\cite{DEGraph}. 
We performed both the signed and unsigned analyses.  
For each of these, we also had to make a decision regarding
how to handle pathways with more than one connected component, and
hence more than one $p$-value.  We tried both simply taking the $p$
value for the largest connected component as the $p$ value
for the pathway, as well as combining the $p$ values for all the
components using  Stouffer's method.  DEGraph will report both the
non-networked $p(T^2)$ as well as the graph--smoothed $p(T^2)$graph;
the latter is the primary quantity of interest. 
\ifTGSA{\item[TopologyGSA]~\\
Finally, we attempted to apply TopologyGSA~\cite{topologyGSA,topologyGSAR}
as implemented to our data.  In principle, TopologyGSA reports $p$
values for both differential variance and differential mean expression
across the pathway submodules, both of which are of interest.
}\fi
\end{description}

\comment{and the ``self--contained'' null is both
better justified biologically (as it doesn't create physiologically
unrealistic pathways) and directly answers the question of whether 
the pathway is associated with the phenotype.  Because the two tests
represent two different conditional probabilities (``competitive''
being conditioned on the sample labels, allowing the definition of the
pathway to vary; the ``self--contained'' being conditioned on the
pathway definition, but allowing the sample classes to vary), they
cannot be directly compared.}

\section{Results}
\subsection{Computational efficiency}
The computational time to complete each  each of the analyses on a
desktop machine (3.4 GHz quad-core Intel Core i7 iMac with 16GB
RAM) is given in Table~\ref{runtab}.
Each of the methods shown in Table~\ref{tab01} was applied to all
10 data sets described in Table~\ref{ovdattab} for 247 pathways.   
Where permutation tests were required, 1000 permutations were used.
All jobs completed the calculation in under an hour per study with
the exception of NEA, which required ${\sim}2.5$~hrs/study\ifTGSA{,
and TopologyGSA, which failed to complete even the first analysis
when it was finally halted after 100hrs (${>}4$~days)}\fi.  
Interestingly, ROT/pe, which is a weighted modification of the same
computation carried out in SPIA, required less time than
SPIA (and, additionally, was able to treat more pathways), which
we attribute to code improvements by the authors of both
methods~\cite{SPIA,SPIAR,draghici2007systems,ROTPE,ROTPER} and made
ROT/pe amongst the fastest of the methods we tested.  Nevertheless,
with most methods taking only a few minutes per study, the differences
in computational cost between them are minor.

With the exception of 
\ifTGSA{TopologyGSA, which we will return to below, and }\fi 
DEGraph, the major computational cost is due to
permutation testing.  (In DEGraph, the computation of the smoothing
vector scales as $\bigO(n^3)$ where $n$ is the number of genes in
the pathway graph, and so can be cumbersome for very large pathways;
however, DEGraph does not require permutation tests.)  Permutation
testing is trivially parallelizable, and the development of parallel
R libraries such as snow~\cite{snowR} facilitates development of
packages that can be run on clusters.  Yet, of the packages considered
here, only CePa provides a parallelized implementation.

\ifTGSA{On inspection, it became clear that TopologyGSA  had been
working for over 90 CPU hours to obtain the maximal cliques for a
single pathway.  As expected, the clique problem had become
intractable; an example of the computational cost of the clique
algorithm used in TopologyGSA for a moderately sized KEGG network
is given in Supplementary Table~\ref{suptab1}.  Unfortunately,
TopologyGSA has no mechanism built in to monitor and bail out of
the computation, nor does it check the size of the transformed
pathway before invoking the max clique algorithm so that potentially
problematic pathways can be skipped.  Running TopologyGSA thus
requires that the user confine the analyses to smaller pathways,
but because the computation depends on the density of the graph
\textit{after} moralization and triangularization, doing so is not
as simple as merely selecting smaller pathways.  While the user
could transform the graphs by hand to make a selection of tractable
pathways, these checks should really be built into the package.
The issue of exponential cost should also be well--documented,
ideally with a small benchmark that the user can run to estimate
the computational time required on their machine (\eg, extrapolating
from a table such as Supplementary Table~\ref{suptab1}), so that
the user can make informed decisions about the application of the
method to particular pathways.}\fi

\subsection{Analyses}
As discussed above, we posit that if a pathway is functionally
related to a particular phenotype (here, high vs. low grade ovarian
cancer), we expect that a manifestation of its involvement will be
present in the data for all studies of that disease, and that an
accurate and sensitive network analysis approach will detect those
signals consistently across the studies while a poor network analysis
method will yield results that are strongly influenced by noise in
the data and will vary strongly from one study to another.  Based
on this intuition, we consider the cross--study concordance of each
method's results to measure its ability to detect a common (and
presumably ``true'') signal in each of the studies.

\paragraph{Cross--study concordance}
For each method tested, we examined the correlation of $p$-values
obtained for the 247 pathways between all 45 possible pairs of the
10 studies listed in Table~\ref{ovdattab}.  For reference, we also
began by examining the correlation in gene--level statistics between
the studies.  Because each method has a different range/resolution
of possible $p$ values owing to the different uses of parametric
and nonparametric tests, we use the nonparametric Spearman's rank
correlation as a measure of concordance.

Figure~\ref{windowPane} depicts the correlation of results for all
all 45 study pairs.  For completeness, we show the cross--study
correlations for each method's subanalyses, in addition to the
``final'' combined result.  This includes the $p$ORA and $p$Pert
values for SPIA and ROT/pe (both with and without a cutoff) that
are combined to form the final $p$Comb value for the pathway; the
standard $p$Hyper hypergeometric test along with the network
edge--based hypergeometric test from PathNet; the results for all
six centrality measures considered by CePa ORA and GSA analyses,
which are then combined into $p$Comb; and the DEGraph results for
the standard Hotelling $T^2$ and the network--smoothed $p$T2graph
for the largest connected component (lcc) and the full pathway
($p$'s for all connected componenents combined) for both the signed
and unsigned analyses.  The ``final'' results for each analysis are
denoted by bolding.  Supplementary Figure File~\ref{xStudyPairs}
provides a more detailed view of the these correlations.  To aid
in the interpretation of Fig.~\ref{windowPane}, Fig.~\ref{xStudyCorl}
presents a summary of the cross--study correlations for each of the
``final'' results.  Here, the distribution of correlations for each
of the 10 studies with respect to all other studies is shown for
each of the methods.

As can be seen in the top row of Fig.~\ref{windowPane} and first
panel of Fig.~\ref{xStudyCorl}, the correlation of gene--level
statistics is often poor; we obtained a maximum rank correlation
$\rho=0.20$ for the gene $p$-values, with a median of $\sim0.02$.
This lack of correspondence even amongst studies of the same phenotype
has been observed in other investigations of microarray data~\cite{MANO06}.
Also as previously observed in other studies~\cite{BRAU2010,MANO06},
we found that the correlations were generally improved in pathway
analyses, both in terms of the number of positively--correlated
study pairs (fewer blue cells in Fig.~\ref{windowPane}) and in the
magnitude of the positive correlations (cf
Figs.~\ref{windowPane},\ref{xStudyCorl}).  Considering that each
study is interrogating gene expression in the same phenotypes, we
ideally desire that the correlations between all study pairs should
be positive, and indeed many were, with the 
CePa GSA analyses being the strongest and most consistent (cf
Fig.~\ref{xStudyCorl}).

In addition, we also observe that the concordance for the network--based
analyses was always stronger than that exhibited by non--network
analyses.  In Fig.~\ref{windowPane}, this can be seen in the darker
blue cells of GSEA and both SPIA and ROT/pe $p$ORA sub-analysis
values, all of which measure enrichment without considering the
network topology.  In Fig.~\ref{xStudyCorl}, this difference is
manifest by the lower medians and tails for the GSEA boxplot versus
the others.

It is also instructive to consider the cross--study correlations
for each method in the context of the gene--level concordance and
the sample size of the studies.  The study pairs depicted in
Fig.~\ref{windowPane} are sorted in order of gene--level concordance.
Generally, we expect that studies which have greater similarity at
the gene level will also exhibit greater similarity in the pathway
statistics, and in most cases this pattern holds, with lower
correlations at the left end of the plot.  Qualitative
departures from this trend can be seen in the SPIA, NEA, and
CePa--ORA betweenness results.  Moreover, we expect that two large
studies will exhibit greater correlation than two smaller studies,
on the basis of the intuition that two large studies are better
powered to distinguish subtle but biologically meaningful, and hence
consistent, effects from noise.  In Fig.~\ref{windowPane}, the sum
of the sample sizes for each study pair is given in the bottom row,
and the pattern of study sizes can be seen to be most strongly
manifest in the various CePa analyses.

\paragraph{Consistency between methods}
A similar intuition regarding the concordance of findings
may be had regarding the results of the various methods
\textit{within} a given study.  That is, despite the differences
between the methods, we might reasonably expect that if a 
pathway is strongly affected in a particular study, it will be
detected by more than one of these methods, \ie, that for 
a given study the results from SPIA, ROT/pe, NEA, \etc will
be correlated.
Figure~\ref{xMethodCorl} shows, for each study, the distribution
of pathway $p$--value correlations that each method had with all
other methods.  A more detailed view is provided in Supplementary
Figure File~\ref{xMethodPairs}.   As above, we use Spearman's rank
correlation as a measure of concordance to account for methodological
differences influencing the dynamic range of $p$ values.  NEA
analyses tended to give results that disagree with methods obtained
from the other analyses (cf Fig.~\ref{xMethodCorl}), for reasons
we ill discuss further below, while the other methods are fairly
comparable.

While the above comparisons consider correlations for the whole
range of $p$ values returned by each method, we now consider how
consistent the selection of ``significant'' pathways is for the
various methods.  The heatmap in Figure~\ref{heatPWoverlap} depicts
the number of times that a pathway was ranked in the top 20\%  for
the 10 studies listed in Table~\ref{ovdattab}.  that a pathway was
ranked in the top 20\%.  Results from all
subanalyses of each method are given, including the $p$ORA and
$p$Pert values for SPIA and ROT/pe (with and without a cutoff) that
are combined to form the final $p$Comb value for the pathway; the
standard $p$Hyper hypergeometric test and the network edge--based
hypergeometric test from PathNet; the results for all six centrality
measures considered by CePa ORA and GSA analyses that are then
combined into $p$Comb; and DEGraph results for the standard Hotelling
$T^2$ test $p$T2 and the network--smoothed $p$T2graph for the largest
connected component (lcc) and the full pathway ($p$'s for all
connected componenents combined) for both the signed and unsigned
analyses.  The ``final'' results for each analysis are denoted by
bolding.  The pathways are sorted by the average across the bolded
analyses, such that (eg) CePa--GSA contributes to that average once
as opposed to seven times.  The probabilities for the counts under
the null hypothesis are also shown; our expectation is that we will
see, by chance, counts of 0--4 with 95\% probability, while counts
$>5$ should occur only once by chance amongst the 247 pathways. On
the other hand, pathways for which the null hypothesis is indeed
false should exhibit high counts more frequently, corresponding to
their detection in multiple studies. Generally, this pattern appears
to qualitatively hold, and the pathways which are deemed significant
in $>5$ studies tend to detected consistently by most of the methods
(with the exception of NEA).  This  suggests that the results are
being driven by commonalities across studies, and that those common
patterns are detectable by many of the methods considered.

\paragraph{Distribution of results}
A more detailed understanding of these patterns emerges from looking
at the concordance between methods for all pathways in all studies,
shown in Fig.~\ref{allCorlPairs}. In the upper triangle of plots,
the joint distributions of $-\log_{10}p$ values are reported; the
corresponding correlation coefficients are shown in the lower
triangle.  Note that, as above, rank correlation coefficients are
reported (and so may differ from ``by eye'' estimates).  On the
diagonal, histograms of $-\log_{10}p$ are plotted in red; the black
lines show the expected distribution of $-\log_{10}p$ corresponding
to the uniform distribution of $p$ values expected under the null.
Strikingly, it can be seen from the histograms in Fig.~\ref{allCorlPairs}
that the $p$ values obtained from NEA 
are strongly biased toward highly significant $p$
values; indeed, over half of all NEA $p$Z  values
fall $\leq10^{-3}$, whereas we expect that the proportion would be
$\sim 1/1000$.  
This causes an extremely large fraction
of pathways to be deemed statistically significant even after
adjusting for multiple hypotheses in NEA, making it difficult to
discriminate truly significant pathways using the current implementation.
Such severely skewed $p$-value distributions are generally attributable
to an incorrect null model.\footnote{\label{NEAnull}In the case of NEA, we believe
there may be a simple remedy for the incorrect null model.
Specifically, we note that the NEA package does not distinguish genes that are
assayed and deemed non-significant from genes those that are simply
not assayed.  By treating non-assayed genes as insignificant, the
proportion of significant genes is significantly reduced,
lowering the probability of significant edges in the resampled
graphs.
\ifarXivHide{We have reported theses issues to the package maintainers
and look forward to the release of corrected code.  We will re-run
our analyses as updates are made, making the results available at
\url{http://braun.tx0.org/netRev}.}\fi}
The other methods follow the theoretical $p$ value
distributions relatively well, with a slight deviation observed
in CePa--GSA.

\section{Discussion} 
Our review indicates a number of benefits and drawbacks associated
with each method, some of which are inherent to the underlying
methodology, and others of which are consequences of the implementations.

\subsection{Methodological considerations}
In Table~\ref{tab01}, the design features of each method are listed.
A major distinction between the methods is the need for gene $p$-value
thresholding.  Threshold--free analyses are generally considered
preferable to those that require gene--level significance thresholds,
since the threshold introduces an arbitrary choice and excludes the
full spectrum of data from the analysis; for this reason, GSEA is
preferred to hypergeometric enrichment tests~\cite{GSEA05,KHAT2012}
for non-network pathway analyses.  Of the network--based methods,
SPIA, PathNet, and NEA have the common drawback of requiring
thresholding on gene--level significance; in contrast, CePa and
ROT/pe provide threshold--free options, and the DEGraph \ifTGSA{and
TopologyGSA }\fi analyses are threshold--free by design.  However,
it should be noted that the cutoff--free ROT/pe analysis differs
substantially from the cutoff based ROT/pe analysis.  Specifically,
ROT/pe with a cutoff computes both the downstream perturbation analysis
($p$Pert) and a hypergeometric over-representation analysis ($p$ORA)
which are then combined to measure the impact of differential expression
on a pathway, while in the cutoff--free analysis, only the 
downstream perturbations $p$Pert are considered.  This limitation could
be overcome simply by using a GSEA-like analysis in place of the
hypergeometric test, enabling both $p$ORA and $p$Pert to be computed 
with or without a threshold.

The second major methodological distinction is in the type of
hypothesis being tested by the methods.  ``Competitive'' null
hypotheses compare the pathway of interest to a random pathway while
holding the sample classes (or gene--level associations) fixed,
whereas ``self-contained'' null hypotheses test if the pathway is
more strongly associated with a particular phenotypic attribute
than expected by chance given the genes and topology of the
pathway~\cite{KHAT2012}.  The two tests represent two different
conditional probabilities (``competitive'' being conditioned on the
sample labels, allowing the definition of the pathway to vary; the
``self--contained'' being conditioned on the pathway definition,
but allowing the sample classes to vary), and may thus give different
results.  The ``self--contained'' null is considered superior since
it is both better justified biologically (``competitive'' permutations
tests will create physiologically unrealistic pathways) and directly
answers the question of whether a particular pathway is associated
with the phenotype of interest.  Unfortunately, though, methods
testing the ``self--contained'' null tend also to be limited in the
type of data that can be used: DEGraph\ifTGSA{, TopologyGSA,}\fi
and CePa-GSA are limited to two--sample comparisons of continuous
data, making them unsuitable for survival analysis or application
to GWAS SNP data.  However, while this limitation is inherent to
the distributional assumptions made in DEGraph (which uses Hotelling's
$T^2$ test) \ifTGSA{and TopologyGSA (which uses a Gaussian network
model)}\fi, it is only an implementation limitation in CePa-GSA
rather than a methodological constraint.  A revision of the CePa
package with a more flexible interface would provide a threshold--free,
``self--contained'' network analysis tool that could be applied to
a broad variety of studies.

\subsection{Ease of use}
All network methods tested are provided as R or BioConductor packages,
making them easily adoptable.  In addition, SPIA, ROT/pe,
\ifTGSA{TopologyGSA,}\fi and DEGraph accept R graphNEL~\cite{Rgraph}
objects describing pathways, making them easy to use with pathway
annotation packages such as KEGGgraph~\cite{KEGGgraph},
GRAPHITE~\cite{Rgraphite}, NCIgraph~\cite{Rncigraph}, \etc, without
additional preprocessing. 

A more serious consideration regarding the ease of use is the
computation time, shown in Table~\ref{runtab}. As discussed above,
most methods are comparably efficient with the exception of NEA
\ifTGSA{and TopologyGSA}\fi.  We could identify no methodological
factor that contributes to NEA's lengthy run-time, and believe that
it is likely due to inefficiencies in the R implementation.  \ifTGSA{In
contrast, TopologyGSA's inefficiency is an unavoidable consequence
of the method's reliance on solving the NP-hard maximal clique
problem.  The exponential complexity can easily become intractable
for even moderately sized pathways, illustrated in Supplementary
Table~\ref{suptab1}.  At minimum, checks should be built into the
TopologyGSA package to conservatively skip pathways that may fail
to be solved in a reasonable amount of time (\eg, by rejecting
pathways with more than a certain number of nodes or edges after
moralization and triangulation).  However, it is not clear that the
maximal clique problem necessarily needs to be solved.  Arguably,
simply detecting community structure within the network (\ie, finding
dense subgraphs without requiring that all pairs of nodes are
completely connected as in a clique) is sufficient to define pathway
``modules,'' which can then be analyzed in the same way as the
cliques currently are.  Detecting community structure/clusters
within a graph is readily achieved by spectral
methods~\cite{CHUN97,NG2002,Newman:2004p2491,Danon:2005p2497,BRAU2010},
that solve a relaxation of the problem in $\bigO(n^3)$ time.}\fi ~Computational 
efficiency of many of these methods could be further improved by
parallelized implementations, a feature that only CePa has to date.

\subsection{Reliability of results}
In absence of ``gold--standard'' data against which to benchmark
these analyses, we attempted to characterize their reliability
based on the concordance of their results in a suite of comparable 
ovarian cancer studies (Table~\ref{ovdattab}).  

Based on the intuition that studies of similar phenotypes should
yield similar results, we examined the correlation of pathway statistics
obtained by each method amongst the 10 studies.
In general, we found greater concordance in the pathway network analysis 
results than in the simple gene--level analysis (cf.
Figs.~\ref{windowPane},\ref{xStudyCorl}), which was the expected
and desirable result from studies of the same phenotype~\cite{MANO06}.
We also observed that the network--based analyses generally gave more
concordant results than the non-network GSEA analysis (Fig.~\ref{xStudyCorl}).
The greatest increases in cross--study concordance were obtained with 
ROT/pe and CePa (Figs.~\ref{windowPane},\ref{xStudyCorl}).  In addition,
we expected that meaningful cross--study concordance of pathway
results would be positively influenced by both gene--level concordance
and the power of the studies under consideration.  With the exceptions
of NEA and the SPIA analyses, the correlation
of pathway--level concordance with gene--level concordance is visible
for all methods in Fig.~\ref{windowPane}, meeting our expectations.
The correlation of concordance with sample size is most clearly
manifest only in the CePa-GSA analyses, and, weakly, in DEGraph.
We are thus inclined to believe that the improved cross--study 
concordance amongst the CePa and DEGraph are attributable to their
detection of common biological signals across the 10 studies.

Based on the conjecture that biological ``truths'' should be
detectable despite slight methodological differences between the 
analyses, we also examined the concordance of findings between
various methods.  Fig.~\ref{xMethodCorl} 
shows the correlations in pathway $p$ values amongst the methods
for each study.  Most methods yielded comparable results, with the
exception of NEA.  Correlations
in the pathways consistently identified as being in the top 20\%
significant can also be seen amongst SPIA, ROT/pe, PathNet, CePa, and DEGraph
in Fig.~\ref{heatPWoverlap}.  By contrast, NEA tends 
to detect consistently significant pathways that are infrequently
found amongst the top 20\% in any study using the other methods.

In addition to investigating the cross--study and cross--method
concordance,  we also examined the distribution of $p$ values
obtained for the methods, shown in Fig.~\ref{allCorlPairs}. It is expected
that most of the 247 pathways tested are \textit{not} significantly
associated with ovarian cancer, and thus the distribution of $p$
values should be, by definition, uniformly distributed on $[0,1]$
with the exception of a handful of significant pathways.  However,
as seen in Fig.~\ref{allCorlPairs} and discussed above, 
the network enrichment $p$Z computed in NEA are exceedingly small
a majority of the time, indicating a likely problem with the null
model used in NEA.\footnotemark[1]

More generally, this observation raises questions
about what constitutes an appropriate null model for network
analyses.  In the context of the ``self--contained'' hypothesis
tests, the answer is straight--forward: one permutes the sample
labels, leaving the network itself intact.  For the ``competitive''
hypothesis, however, the answer is far less clear.  The difficulties
in constructing null models that accurately preserve the statistical
and graph theoretic properties of networks have been considered by
ourselves and others~\cite{BRAU2010,LEIB08}.  Most notably, simple
randomization of node properties or graph rewiring will produce
null models that lack the assortativity found in biological interaction
networks.  That is, because of the underlying biology, groups of
genes that are connected in pathway databases will likely exhibit
correlated expression (and hence correlated gene--level significance)
in experimental data.  By randomizing the data across the network,
that property is destroyed, resulting in network models that are not
as structured as those found in nature.  Methods using such null models
will yield inflated significance, since the data is being compared
against \textit{naive} random networks rather than \textit{biologically
plausible random networks}.  The test of the ``competitive'' null
hypothesis thus demands more sophisticated null models than those
currently employed.  These issues also underscore the benefit of
using a ``self--contained'' test, which preserves the associations
between gene expression and network structure  while randomizing
their association with the phenotype of interest.

In general, we find that CePa-GSA exhibits the best cross--study
concordance (Fig.~\ref{xStudyCorl}), does so in a way that meets
reasonable expectations of being correlated with gene--level
concordance and sample sizes (Fig.~\ref{windowPane}), and has the
benefit of testing the prefered ``self--contained'' null.  The
drawbacks of the method, however, include limitations regarding the
input data as discussed above, as well as the fact that CePa returns
several sub--analyses that must be selected or combined by the user,
as we did here.  CePa-GSA also required nearly an hour per study
when using 1000 permutations, though this may be possible to reduce
with CePa's parallel implementation.  Other methods provide speedier
computations and more flexible inputs, albeit at the expense of
improved concordance or other methodological limitations
(cf.~Table~\ref{tab01})

\section{Conclusions}

New network--based methods have garnered increasing interest as
tools to analyze complex genomic datasets at the systems level.
Despite the development of a number of promising tools, however,
there is little guidance available to researchers for choosing 
between the methods.  In this review, we sought
to compare all the network analysis methods available in R/BioConductor
at the time of this writing~\citeAll.  In addition to discussing
their methodological and implementation features, we also proposed
and applied a novel means to compare their performance using a suite
of curated microarray data-sets~\cite{GANZ2013} and a set of updated
KEGG mappings developed to enable consistent pathway models for
each method.  The data we used for the analysis, the prepared KEGG
pathways, and the scripts to carry out the computations (including
functions to refresh the curated data and KEGG mappings from their
source repositories) have been made available on the author's website
(\url{http://braun.tx0.org/netRev}) to enable other researchers to
apply these comparisons to new methods as they are developed.  In
addition, we plan to make available updated versions of our findings
as these packages are updated.  The results of our tests clearly
indicated the benefits and limitations of each approach.  The tests
also revealed idiosyncracies that would have been unnoticed except
in comparison; for example, our comparisons revealed a bug in the
previous version of the ROT/pe computation, which led us to suggest
a fix that has now been implemented in the current version (reviewed
here).

In addition to providing guidance about the features of the methods
(Table~\ref{tab01}), the efficiency of the computations
(Table~\ref{runtab}), and the consistency of the results
(Figs.~\ref{windowPane}--\ref{allCorlPairs}), our review also suggests
a number of directions for future methodological development. 

Most notably, there is a need for benchmark and testing standards
against which network analysis methods should be tested.  We used
the consistency of the results across a set of comparable studies,
but this approach is plagued by a serious limitation: namely, we
have no way to assess whether the ``consistent'' results are consistent
owing to biological commonalities amongst ovarian cancers or due
to a fluke of the microarray data, since the set is homogenous 
with respect to the disease type.  A more insightful analysis could
be obtained by the development of a database of diverse studies
that are all curated to the same standards, just as was done for
the curated ovarian data~\cite{GANZ2013}.  While diverse datasets
are readily obtained, the work required to ensure that they are
all comparable is non-trivial (and was beyond the scope of this 
paper); however, such data would be immensely useful to the 
research community.   Relatedly, agreement on a common pathway
representation format such as BioPAX~\cite{BioPAX} and developers'
adoption of a consistent API accepting these pathway files would
aid comparison between these methods without requiring that the
pathways be prepared by the user in different ways.

Secondly, we note that the most significant methodological distinctions
between the packages involve a choice between using the preferred
``self--contained'' null hypothesis versus having the flexibility
to apply the method in contexts other than two--sample differential
expression studies.  We recommend using methods that test the
self--contained null (both for statistical and biological reasons
as discussed above and in~\cite{KHAT2012}), but at present none of
these packages are able to test, for example, a self--contained
hypothesis that a pathway is significantly associated with survival.
This compromise could be easily resolved by further development of
CePa--GSA allowing the user indicates to the function the statistical
test (or model to be fit) rather than assuming that a two--sample
$t$-test is desired.  In the case of DEGraph, which uses Hotelling's
two-sample $T^2$ statistic to compare the graph ``smoothed'' gene
expressions in two phenotypes, such an extension is less obvious but
would be a valuable addition to DEGraph's functionality.

Relatedly, we note that care must be exercised when constructing
null models for the pathways for the ``competitive'' tests.  An
easy check of whether or not the null model is correct is to examine
the distribution of $p$ values across a large set of pathways;
strong deviation from the expected uniform distribution of $p$
values is indicative of an incorrect null model.  However, this
rough assessment will only reveal egregious flaws.  In the methods
discussed above, and indeed many network biology methods generally,
null graphs are generated by simply resampling node or edge properties.
This destroys the correlation structure in the data (as \cite{GSEA05}
discussed) as well as the expected assortativity of gene expression
in the pathway, yielding excessively conservative null models. There
is thus a need to develop methods that can produce null graphs that
are more biologically plausible.

Finally, we observe that a common drawback to all of these methods
is their reliance upon single--gene statistical tests.  As a result,
while all of these methods are able to articulate differences in
gene expression that have an impact at the pathway level, they
cannot detect differentially regulated pathways when there are no
detectable marginal effects at the gene level. An alternative
approach would be to overlay the gene expression data itself onto
the network (instead of using statistics corresponding to the gene's
differential expression), obtain a summary statistic for the
network as a whole, and compare those.  This approach has proved
powerful in a non--network context~\cite{BRAU2008,BRAU2010}, where
it was able to detect pathways in which non-linear patterns of
gene expression were associated with phenotype.  While network
extensions have been proposed~\cite{BRAU2012}, R implementations
remain lacking.  

Network analysis is rapidly becoming a valuable tool for harnessing
existing biological information to yield mechanistic, systems--level
insights from HT data.  A number of promising methods have been
developed, and we have found that most yield more consistent results
(as measured by cross--study concordance) than both gene--level
analyses and non--network pathway analyses (GSEA~\cite{GSEA05}).
Nevertheless, challenges remain, and further work in this area has
the potential to significantly improve the systems--based analysis
of HT data, facilitating better understanding of the structure and
function of the complex networks that coordinate living processes.

\clearpage
\clearpage

\methTab

\studTab

\timeTab

\clearpage

\figsection 

\keggfigFig

\windowPaneFig

\xStudyCorlFig

\xMethodCorlFig

\heatPWoverlapFig

\allCorlPairsFig

\clearpage
\bibliographystyle{unsrt}
\bibliography{journals_short,allBib3}
\clearpage
\startsup
\section{Supplementary Information}
\ifTGSA
\begin{table}[!h]
\centering
\begin{tabular}{p{6in}}
\begin{lstlisting}
> hsa00190.gnl
A graphNEL graph with directed edges
Number of Nodes = 133 
Number of Edges = 132 

> (hsa00190.mrl <- moralize(hsa00190.gnl))
A graphNEL graph with undirected edges
Number of Nodes = 133 
Number of Edges = 1078 

> qpGetCliques(hsa00190.mrl)
  1/133 (max 1)      0.00 s      (0.00 s/round)
104/133 (max 1)      0.10 s      (0.00 s/round)
105/133 (max 1)      0.20 s      (0.10 s/round)
106/133 (max 1)      0.41 s      (0.20 s/round)
107/133 (max 1)      0.82 s      (0.41 s/round)
108/133 (max 1)      1.66 s      (0.84 s/round)
109/133 (max 1)      3.32 s      (1.66 s/round)
110/133 (max 1)      6.66 s      (3.34 s/round)
111/133 (max 1)     13.35 s      (6.70 s/round)
112/133 (max 1)     26.78 s     (13.42 s/round)
113/133 (max 1)     53.70 s     (26.92 s/round)
114/133 (max 1)    107.67 s     (53.98 s/round)
115/133 (max 1)    215.86 s    (108.19 s/round)
116/133 (max 1)    432.71 s    (216.85 s/round)
117/133 (max 1)    866.78 s    (434.07 s/round)
118/133 (max 1)   1736.20 s    (869.42 s/round)
119/133 (max 1)   3477.42 s   (1741.23 s/round)
120/133 (max 1)   6964.69 s   (3487.27 s/round)
121/133 (max 1)  13948.89 s   (6984.19 s/round)
122/133 (max 1)  27948.81 s  (13999.92 s/round)
...
\end{lstlisting}
\end{tabular}
\caption{\label{suptab1} R output showing exponential cost of the maximal clique
problem for a the oxidative phosphorylation KEGG pathway. hsa00190
is a modestly sized 133 nodes and 132 edges (below
the median across all KEGG graphs). After moralization, however,
it becomes considerably more dense, jumping to 1078 edges.  Finding
cliques in a graph of this size is a challenging task.
TopologyGSA does this with \texttt{qpGetCliques()}, which is an
R interface to the GNU GPL Cliquer library implementing {\"O}sterg{\aa}rd's
algorithm~\cite{ostergaard2002fast}, the fastest maximal clique
algorithm to date.  The algorithm uses a branch-and-bound procedure,
in which first small subgraphs of $S$ nodes are searched for maximal
cliques.  Once found, the size of the subgraph is incremented and
searched again, until either all nodes are in the subgraph $S=133$
or when increasing the subgraph would not permit the maximal clique
found to that point.  Because maximal clique is an NP-hard problem,
the cost increases exponentially with each increase in subgraph
size, as can be seen here with the dense moralized hsa00190 graph.
The first column of the timing output shows the progress of the
calculation as the subgraph size is increased.  In the second column,
the accumulated runtime is recorded, while the last column gives
the amount of time required since the previous search round finished
clique was found.  The exponential increase in hardness as is clearly
visible.  Timing was stopped after approximately 8h, before the
algorithm could complete.  Extrapolating from this data, running
to completion could take as much as ${\sim}57042500$sec, nearly two years.}
\end{table}
\clearpage
\fi

\paragraph*{Figures \ref{xStudyPairs}: supp.xStudyPairs.pdf}
\refstepcounter{figure} \label{xStudyPairs} ~\\
{Plots of $p$-value correlations between different studies for
each of the 10 methods, along with the gene-wise $p$-value 
correlations.  $-\log_{10}p$ values are plotted against
each other above the diagonal; below, the correlation coefficients
are given.  Note that in NEA (pg 9), a
huge number of pathways ``saturated'' the test, with none of 1000 
permutations yielding statistics comparable to those observed,
suggesting possible issues with the null model.}

\paragraph*{Figures \ref{xMethodPairs}: supp.xMethodPairs.pdf}
\refstepcounter{figure} \label{xMethodPairs} ~\\
{Plots of $p$-value correlations between different methods for
each of the 10 studies.  $-\log_{10}p$ values are plotted against
each other above the diagonal; below, the correlation coefficients
are given.}

\end{document}